# Using Data Analytics to predict students' score

**MA Nang Laik\*, Chua Gim Hong**
School of Business
Singapore University of Social Sciences
Singapore
nlma@suss.edu.sg

**Abstract**

Education is very important to Singapore, and the government has continued to invest heavily in our education system to become one of the world-class systems today. A strong foundation of Science, Technology, Engineering, and Mathematics (STEM) was what underpinned Singapore's development over the past 50 years. PISA is a triennial international survey that evaluates education systems worldwide by testing the skills and knowledge of 15-year-old students who are nearing the end of compulsory education. In this paper, the authors used the PISA data from 2012 and 2015 and developed machine learning techniques to predictive the students' scores and understand the inter-relationships among social, economic, and education factors. The insights gained would be useful to have fresh perspectives on education, useful for policy formulation.

## 1. Introduction

In Singapore, the government has invested heavily in our education system into a world-class system today. A strong foundation of Science, Technology, Engineering, and Mathematics (STEM) was what underpinned Singapore's development over the past 50 years. STEM education in Singapore integrates the four disciplines into a cohesive learning paradigm based on real-world applications. There is a demand for STEM-related jobs across the globe, with even the United States, United Kingdom, and Germany facing huge employment shortage in the STEM field. According to the US Department of Statistics, STEM occupations are growing at 17% while other occupations are growing at 9.8%.

There has been an abundance of studies on the factors affecting academic performance in subjects such as Reading, Science, and Maths. However, they have focused on students' performance in the U.S., Europe, and countries other than Singapore. They also tended to focus on undergraduates and college students. Since there were cultural differences between these countries and Singapore, and furthermore, such differences might play a role in shaping the factors that affected academic performance, it would be very important to examine those factors which were relevant to Singapore students' performance in Reading, Maths, Science and Problem Solving. The data of Singapore students who participated in the OECD Programme for International Student Assessment (PISA) provided this opportunity, as it remains one of the most comprehensive data ever collected.

PISA is a triennial international survey which evaluates education systems worldwide by testing the skills and knowledge of 15-year-old students who are nearing the end of compulsory education.

Over 90 countries have participated in PISA which began in 2000. Every three years, students are tested in Reading, Mathematics, and Science, and an innovative domain such as problem-solving in 2012 and 2015. Not only do the students take a test, but they also fill out a background questionnaire to provide contextual information e.g. they are asked about the level of education of their parents and possessions in the household e.g. computers, air conditioning, desk, car, books, bath/shower rooms, phones, literature, software, etc. Also, school principals fill out a questionnaire about how their schools are managed. There are also options to administer (1) Student's Questionnaires on Financial Literacy, Educational Careers, and Use of ICT, (2) Parent's Questionnaires, and (3) Teacher's Questionnaires.

Previous literature also did not suggest the use of data analytics to predict the students' score. It will be one of the innovative way to look at PISA data for Singapore. The authors would like to develop a data mining or machine learning model to predict the academic performance of the students based on factors that were student-related, school/class related, teacher-related, home-related, and parent-related. Analytical models could help us to plan and implement subject-based banding (SBB) and STEM policies by helping us to understand the factors which affect academic performances in STEM with social and psychological factors.

The rest of the paper is organized as followed. Section 2, there will be a literature review to look at the related research which looks at students' academic performance from various countries using qualitative and quantitative approach. Section 3, exploratory data analysis will be performed and the authors would share some interesting insights. In section 4, data mining models will be developed and computational results will be shown. Finally, future research directions and implementation issues will be discussed.

## 2. Literature Review

There have been many studies that sought to understand the influences of key factors on students' academic performance. Most of those studies have focused on students' performance in the U.S. and Europe, with some studies from other countries. These factors could be broadly classified into student-related, teacher-related, school-related, parent-related, and home-related, including socio-economic, psychological and environmental factors such as family background, students' characteristic (e.g. self-efficacy, learning style, study habits, psychology), school experience factors, student attendance, extra-curricular activities, peer influence, hard work, discipline, previous schooling, parents' education, family income, self-motivation or even gender (Eccles & Harold, 1993).

(Chen & Lin, 2006) found that class attendance has produced a significant positive impact on students' exam performance. (Arulampalam et al., 2008) found that absenteeism led to poorer performance. Some researchers found no significant relationship between class attendance and academic performance (Martins and Walker, 2006; Caviglia-Harris, 2006).

Various studies have found that peers influence was positively related to student examination performance (Goethals, 2001; Gonzales et. al., 1996; Hanushek et. al, 2003). (Wilkinson and Fung, 2002) found that grouping students in heterogeneous learning ability (low ability students grouped with high ability students) resulted in improvement in the learning process and outcomes. In another study, (Schindler, 2003) found that mixing students of different abilities in the same class would affect weak students positively however the effect for good students was negative. This was in contrast with (Goethals, 2001) who found that students in a homogeneous group, whether high or low ability, performed better than students in a heterogeneous group. (Lujan & DiCarlo 2005) found that peer interaction could increase students' participation and ability to problem-solving.

(Romer, 1993) found that class attendance significantly impacted students' academic performance. (Ellis et al., 1998) found that absenteeism negatively affected student performance in principles of economics. (Tay, 1994) concluded in their survey research on the factors affecting students' performance in economics that students' aptitude was the most important determinant of learning. Study effort, age of the student, and a good match between students' learning style and instructor's teaching style all have a positive effect on students' performance.

Using regression model and a sample of 864 students from the College of Business and Economics-UAEU, (Nasri & Ahmed, 2006) found that the most important factor affecting students' performance was students' competence in the language medium used to teach the subjects i.e. English. Besides competence in English, students who participated in class discussions outperformed other students, non-national students outperformed national students and female students outperformed male students. The factors that most negatively affect students' performance were missing too many lectures and living in a crowded household.

Using K–8 national longitudinal data, (Robinson & Lubienski, 2011) found that preschool and primary school boys and girls generally performed similarly on Maths tests, except among the higher-performing students where males performed better in Maths than females. Males performed better than females in Maths in high school and college, with this difference being larger among higher-performing students but not necessarily for lower or average-performing ones. Moreover, males outperformed females on Maths tests that were less related to what was taught in schools (e.g. SAT math test). There were minimal gender differences on state-wide standards-based Maths tests, which were more tied to what was taught in schools. When it came to school's grades, which were even more closely tied to the curriculum, girls often outperformed boys. Overall there were only small differences in boys' and girls' Maths performance; those differences depended on the age and skill level of the students, what type of Maths they were attempting and how big of dissimilarity was needed to say that boys' and girls' Maths performances were truly different. The authors found females performed better than males in Reading, and this gap generally widened among low-achieving students.

In a study of 114 children with roughly average IQs and typical development and just over half girls, neuroscientists (Escovar et al., 2016) at Stanford University found that children aged 7 to 12 who were ranked higher on their empathetic dispositions performed worse in Maths problems like subtraction, multiplication, or geometry. While high empathy was correlated with lower Maths scores, Reading was not. That, the authors wrote, pointed "against a broad effect of empathizing leading to divided attention in classrooms and suggesting instead that sensitivity to emotional states might be particularly detrimental during Maths instruction."

From the literature review, the authors have identified that there are very few researchers who look at PISA data to predict the students' scores for Singapore students. Thus, the main aim of the paper is to use PISA data from 2012 and 2015 to better understand the students' academic performance in STEM subjects and use data analytics to predict the students' scores. Furthermore, the factors that are important will be identified and will be used to fine-tune Singapore's education system in the future.

## 3. Problem Description and EDA

The Ministry of Education (MOE) in Singapore will replace the Normal (Technical), Normal (Academic), and Express streaming in secondary schools with full subject-based banding (SBB), starting in about 25 schools in 2020, before rolling it out to all secondary schools by 2024. Tailoring subjects for students based on their academic abilities and organizing the school's curriculum to meet the students' specific learning needs and abilities require extensive planning and resources. The classes, which mix students from all three streams, are configured according to SBB for academic subjects such as English, Maths, and Science, as well as to provide the opportunity for students with different academic abilities and from different social-economic background to interact. It would be useful to have Analytical tools that can help us understand the correlation and dependency of a student's performance and abilities in different subjects, as well as with his/her characteristics, social-economic demography, school, or class management, and environment, parent- or home-related factors.

Studies have suggested that about 5% to 12% of the population has dyslexia, where Reading is fraught with difficulty and spelling is sometimes a problem, too. An unknown percentage of the population also grapples with so-called dysgraphia, an inability to write. These often affected their learning of Science and Maths. On the other hand, about 6% to 8% of the world population has dyscalculia or difficulty not with Reading and writing, but with learning basic Maths. It would be useful to have Analytical tools that can help provide an initial screening of students with potential dyslexia, dysgraphia, and dyscalculia, to provide the education support to prepare them for the STEM future.

Singapore has sent students to participate in the PISA's cognitive assessments in Reading, Maths, Science, and Problem-solving using computer-based assessment. PISA also includes students' background survey, and students' use of ICT as well as school principal's survey on school management. The datasets of Singapore 2012 and 2015 PISA scores are available at the point of the research conducted in 2019. The PISA dataset is structurally challenging to analyze and requires a significant amount of time to do the data preparation using advanced techniques and mapping. By learning from the data preparation and cleaning, analysis, and models built for the 2012 and 2015 dataset, this paper aims to provide users with analytical approaches and tools to analyze similar data for future analysis.

2012 and 2015 data include 5546 Singaporean students and 6115 Singapore students record. In 2012 data, there are both paper-based assessment (PBA) and computed-based assessment (CBA) results, however, 2015 data only includes the computer-based assessment result. Data are clean and converted into the required format for analysis. The credit scores are also normalized and full credit is counted as 1, partial credit is 0.5, and 0 for no credit for each subject. The authors then derived the aggregate score for each student for all subjects using the formula:

Sum of scores = (sum of full credit * 1) + (sum of partial credit * 0.5)
Total credit taken = sum of full credit + sum of partial credit + sum of no credit
Aggregate Scores = (Sum of scores) / (total credit taken)*100%.

There are 200 over variables and a lot of missing data are found. Student ID is used to merge the record for analysis purposes. In 2012, 43 countries participated in the exams and the total number of students is 271,323, and Singapore students contributed to 2%. In 2015, the number of countries has increased almost double to 73 and the total students' number has increased by almost twice to 519,334 students. 95% of the students from Singapore are from public secondary schools and only less than 5% were from the private schools. The proportion of male and female students are equal. The following acronyms were used. DRA (Digital reading assessment), CBAM (Computer-based assessment maths), CBAPS (Computer-based assessment problem solving), CPS (Collaborative problem-solving.

Table 1: Comparison of 2012 paper-based and computer-based PISA results with 2015 computer-based Results

| S/N | 2012 (Mean, Std Dev) | 2015 (Mean, Std Dev) | Comparison of Mean |
|---|---|---|---|
| 1 | DRA (73.5%, 20.9%) | Reading (66.6%, 20.0%) | 2012 is higher |
| 2 | CBAM (51.7%, 23.7%) | Maths (60.8%, 22.1%) | 2015 is higher |
| 3 | CBAPS 63.4%, 22.0%) | CPS (64.8%, 16.2%) | 2015 is higher |
| 4 | PBA Read (66.8%, 21.2%) | Read (66.6%, 20.0%) | 2012 is higher |
| 5 | PBA Maths (59.2%, 22.2%) | Maths (60.8%, 22.1%) | 2015 is higher |
| 6 | PBA Science (62%, 22.1%) | Science (58.5%, 21.6%) | 2012 is higher |

Table 1 shows the comparison of results between 2012 and 2015. The mean score for maths and problem-solving are higher in 2015, whereas reading and sciences are higher in 2012. The authors also performed a one-way ANOVA on the mean of the male and female scores and found that there was a significant difference in the mean score of females and males for all the subjects in 2012 and 2015, except for 2012 PBA Science, 2012 CBAM, 2012 CBAPS, and 2015 Maths.

Table 2 shows the correlation matrix between the two subjects. Female's scores are significantly higher than male's scores in sciences and reading. To perform correlation analysis, all possible permutation of column pairs of percentage scores for 2012 and 2015 PISA Reading, Maths, Science, and Problem-solving assessments of each Student IDs were joined in a new table. For each newly generated table with pairs of assessment scores, rows of Student IDs with one or both missing subject scores were filtered away. Correlation analysis was performed on the Mean scores of Student IDs who participated in both subjects were under analysis.

Table 2: Correlation Analysis between two subjects

| S/N | CORRELATION, R | | WHICH PAIR OF PISA SCORES? |
|---|---|---|---|
| 1 | Very High | > 70% | 2012 PBA Reading and 2012 PBA Maths |
| | | | 2012 PBA Reading and 2012 PBA Science |
| | | | 2012 PBA Maths and 2012 PBA Science |
| | | | 2012 PBA Maths and 2012 CBAM |

| S/N | CORRELATION, R | | WHICH PAIR OF PISA SCORES? |
|---|---|---|---|
| | | | 2015 Reading and 2015 Science |
| | | | 2015 Maths and 2015 Science |
| 2 | High | 60% < R ≤70% | 2012 PBA Reading and 2012 DRA |
| | | | 2012 PBA Reading and 2012 CBA |
| | | | 2012 PBA Maths and 2012 DRA |
| | | | 2012 PBA Maths and 2012 CBAPS |
| | | | 2012 PBA Science and 2012 DRA |
| | | | 2015 Reading and 2015 Maths |
| | | | 2015 Science and 2015 CPS |
| 3 | Moderate | 50% < R ≤ 60% | 2012 PBA Reading and 2012 CBAPS |
| | | | 2012 PBA Science and 2012 CBAPS |
| | | | 2012 DRA and 2012 CBAM |
| | | | 2012 DRA and 2012 CBAPS |
| | | | 2012 CBAM and 2012 CBAPS |
| | | | 2015 Reading and 2015 CPS |
| | | | 2015 Maths and 2015 CPS. |

The graphical plots of Reading with Maths, Sciences, and Problem-Solving would represent students who were further away from the prediction or regression line, indicating possible dyslexia (i.e. low Reading and possibly Maths and Science scores), dysgraphia (i.e. low Science and possibly Maths scores) and/or dyscalculia (i.e. low Maths scores). The relatively low $R^2$ values suggest that they might be valuables not in the model that account for the unexplained variation. In the next section, the authors would develop the advanced predictive analytics model using a decision tree, neural network, and multiple regression analysis and share the computational results.

## 4. Predictive modelling

### 4.1 Data Preparation

Data has been prepared using Python and SAS Enterprise miner is used to develop the predictive models to predict the students' scores using the dependent variables such as aggregate score, reading score, math score, science score, and problem-solving score based on the 96 independent student variables. There are some Common variables in the 2012 and 2015 PISA datasets that were deliberately selected for these 96 *Student Variables*. Similarly, predictive models were built to predict the schools' scores (i.e. dependent variables - Aggregate Scores, Reading Scores, Maths Scores, Science Scores, and Problem-Solving Scores) based on the 81 School Variables as inputs. Common variables in the 2012 and 2015 PISA datasets were deliberately selected for these 81 School Variables.

The 2012 PISA datasets were used as the training and validation data for building the Predictive Models, which were then scored on the 2015 dataset. The students in the 2015 dataset were different and about 3 years older than those students in the 2012 dataset. The Average Absolute Percentage Errors between the various predicted and actual 2015 PISA scores (e.g. Aggregate Scores, Reading Scores, Maths Scores, Science Scores, and Problem-Solving Scores) were then computed to measure the accuracy of the predictions. The analytical methods used for developing the Predictive Models were Multiple Regression, Decision Tree, and Neural Network. Since the Target Variable is Continuous or Interval in nature, the Average Squared Error (i.e. average of the square of the difference) between the predicted outcome and the actual outcome was used as the measure for selecting the best model.

Figure 1 shows the process flow diagram for building the predictive model in SAS Enterprise Miner where the Target Variable was students' "Aggregate Scores" and Input Variables were 96 Student Variables. These variables were found in the [2012 Aggregate & Student] "Raw" data node the data was partitioned using 80% training and 20% validation. Three data mining models such as regression, decision tree and neural networks were developed to predict the students' score based on the input. The best model is chosen based on the average square error performance. After the best model is selected, [2015 Aggregate & Student] "Score" data has been scored. The same process flow diagram was also used for different subjects such as "Reading", "Maths Scores", "Science Scores" and "Problem Solving Scores", all of which were computer-based except for 2012 paper-based science.

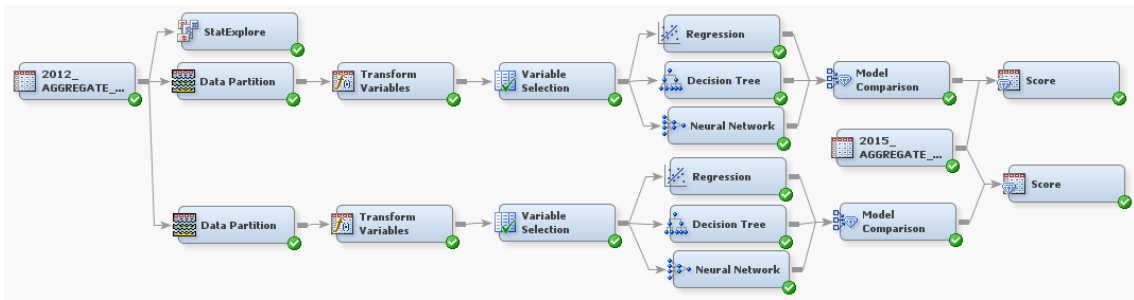

Figure 1: Process Flow Diagram of Predictive Model of Students' Aggregate Scores based on 96 Student variables

The model performance is stated in Table 3 below. Decision tree model is the best champion model as it has the lowest average square error for the validation dataset.

Table 3: Data mining models to predict students score using student demographic information

| Selected Model | Model description | Selection criteria: Validation: Average Square Error |
| --- | --- | --- |
| Yes | Decision Tree | 246.09 |
|  | Regression | 291.74 |
|  | Neural Network | 295.60 |

From the decision tree model, the "Aggregate Scores" of students increased directly with "Science Learning Minutes Per Week" which was something expected. The former also increased directly with the availability of "Literature at Home", higher "Index of Economic, Social & Cultural Status", greater "Number of Computers", more frequent "Use Printer at Home", a higher level of "Repeat ISCED". There were also inverse relationships between "Aggregate Scores" with "Use USB in School", "No. of Books" and "Out-of-School - Upload Content". However, "Aggregate Scores" of students decreased with an increased in "English Learning Minutes Per Week" which was unexpected. Intervening factors probably interacted with antecedents or other intervening factors to produce an interactive effect e.g. if the total amount of learning minutes for all subjects were fixed, then any increase in "English Learning Minutes Per Week" could have perhaps affected the learning minutes per week of other subjects, and hence the performances of those subjects.

The "Aggregate Scores" now also increased as a new Input Variable "Use of ICT for School tasks at Home" increased.

### 4.2 Predictive models of PISA school scores using school variables

To build a model to predict the schools' aggregate scores and various subject scores based on the chosen 81 School Variables. Data is also partitioned with the setting of 80% training and 20% validation, which yields the best result. Some of the variables that are important as the school variables are class size, the number of full-time teachers, student-teacher ratio and offers chess club. Three data mining methods namely decision tree, multiple linear regression and neural network have been used to predict the PISA school score. Table 4 shows the data mining model results to predict school score using school variables.

Table 4: Data mining models to predict school score using school variables

| Selected Model | Model description | Selection criteria: Validation: Average Square Error |
|---|---|---|
| Yes | Decision Tree | 137.45 |
| | Regression | 192.46 |
| | Neural Network | 740.49 |

Decision tree is the best model in this case as it has the lowest average square error of 137.45. Using the decision tree model, the authors can see that the "Aggregate Scores" of students increased directly with "No. of Full-Time Teachers" increased. It also increased with "No. of Boys Enrolled" and when the school "Offers Chess club" and the average predicted aggregate score is 83.3.

The authors also use the formula below to computer the Mean Absolute Percentage Error (MAPE):

=IF(Aggregate Scores > 0, ABS(Prediction for Aggregate Scores - Aggregate Scores) / Prediction for Aggregate Scores, "") * 100%

The summary of Average Absolute Percentage Errors (MAPE) between the schools' "Aggregate Scores" and "Predicted Aggregate Scores" is as shown in Table 5, where the Predictive Model for Maths scores was still the least accurate model. The MAPE for aggregate scores is 12.4% which is considered quite useful and the model is most accurate to predict the problem solving score where the MAPE is 10.6%.

Table 5: Summary MAPE for actual and predicted school scores using school variables

| Mean Absolute Percentage Errors (MAPE) using Predicted vs Actual | 2012 No. of Data Records | 2015 No. of Data Records | MAPE |
|---|---|---|---|
| Aggregate Scores | 172 | 177 | 12.4% |
| Reading Scores | 172 | 177 | 13.7% |
| Maths Scores | 172 | 177 | 25.1% |
| Science Scores | 172 | 177 | 13.4% |
| Problem-Solving Scores | 172 | 177 | 10.6% |

The authors also found out that the School Variables were more accurate than Student Variables in predicting Aggregate, Reading, Maths, Science and Problem-solving scores than using the individual student's variable.

## 5. Moving forwards

The predictive models have been built on the PISA datasets from Singapore students who were born in 1996 for 2012 PISA and 1999 for the 2015 dataset i.e. 16-year old students who were in Secondary 4. The model could be further tested or scored on the 2018 PISA datasets which would be available towards the end of 2019. Note that during the data exploration, it was found that there was negligible or no correlation between Aggregate or various subject scores with the months of birth among these 16-year old students. To operationalize the model, the following steps could be taken into consideration in future.

(1) Conduct surveys for the 16-year old Singapore students (who did not participate in PISA) using the 96 Student Variables (or at least the top 15-30 with the highest variable worth values). Then use the model to predict their "PISA" scores in Reading, Maths, Science, and Problem-solving as well as Aggregate scores. To determine the efficacy of the model, compare these predicted PISA scores with the following to see if the model can accurately predict these results: (1) school curriculum-based exams/tests scores for Reading, Science and Maths and/or (2) school preliminary examination which prepares students for G.C.E. 'O' Levels and/or (3) actual G.C.E. 'O' Level results.

(2) Work with OECD during PISA 2021 to standardize the student variables and school variables and then use the inputs of the student and school surveys to predict the "PISA" scores, and test this prediction against the actual PISA 2021 scores.

(3) If the identity of the students who took PISA scores are known, the authors can compare their PISA scores with their actual curriculum-based exams/tests for Reading, Maths, and Science scores in the school. This could also enable us to determine how accurate are the model developed using actual PISA scores in predicting the results of curriculum-based assessments.

To operationalize the model outside this age range, additional data would have to be collected for the age group of interest. The resource demand would be in designing the assessment questions, administering, marking, and collecting data for the results. The survey on student questionnaires could be based on at least the top 20-30 student variables with the highest variable worth, although there would not be much additional cost to collect all 96 Student Variables in the same survey. This model built in non-PISA settings would then need to be scored against the 2012 PISA (or 2015 or 2018) dataset which would be based on 16-year olds.

To conduct further studies to understand why the student variables and school variables with high variable worth values were able to predict the PISA scores. This could surface factors and reasons previously known, unknown unknowns, or known unknowns that would help in policy formulation to uplift and support students who were not academically inclined.